\documentclass[letterpaper, 11pt]{article}

\usepackage[utf8]{inputenc}
\usepackage[T1]{fontenc}
\usepackage[letterpaper,left=1in,right=1in,top=1in,bottom=1in]{geometry}

\usepackage[utf8]{inputenc} 
\usepackage[T1]{fontenc}    
\usepackage{hyperref}       
\usepackage{url}            
\usepackage{booktabs}       
\usepackage{amsfonts}       
\usepackage{nicefrac}       
\usepackage{microtype}      
\usepackage{xcolor}         
\usepackage{authblk}

\usepackage{amsmath}
\usepackage{graphicx}
\usepackage{caption}
\usepackage{subcaption}
\usepackage{multicol}
\usepackage{float}

\title{De-noising non-Gaussian fields in cosmology\\with normalizing flows}
\author{Adam Rouhiainen\thanks{Electronic address: rouhiainen@wisc.edu}}
\author{Moritz M\"{u}nchmeyer}

\affil{Department of Physics, University of Wisconsin-Madison, Madison, WI 53706, USA}

\begin{document}

\maketitle

\begin{abstract}
Fields in cosmology, such as the matter distribution, are observed by experiments up to experimental noise. The first step in cosmological data analysis is usually to de-noise the observed field using an analytic or simulation driven prior. On large enough scales, such fields are Gaussian, and the de-noising step is known as Wiener filtering. However, on smaller scales probed by upcoming experiments, a Gaussian prior is substantially sub-optimal because the true field distribution is very non-Gaussian. Using normalizing flows, it is possible to learn the non-Gaussian prior from simulations (or from more high-resolution observations), and use this knowledge to de-noise the data more effectively. We show that we can train a flow to represent the matter distribution of the universe, and evaluate how much signal-to-noise can be gained as a function of the experimental noise under idealized conditions. We also introduce a patching method to reconstruct fields on arbitrarily large images by dividing them up into small maps (where we reconstruct non-Gaussian features), and patching the small posterior maps together on large scales (where the field is Gaussian).
\end{abstract}

\section{Introduction}
\label{introduction}

Normalizing flows \cite{2019arXiv191202762P} have been shown to be very effective at learning high-dimensional probability distribution functions (PDFs), in particular when the random variables are spatially organized as in an image. This has led to a lot of recent work where PDFs in physics have been parametrized with flows, in particular in the domain of lattice QCD \cite{Albergo:2021vyo}. In our precurser work \cite{rouhiainen2021}, we evaluated how well various flows can learn sample generation and density estimation of cosmological fields. In the present work, we use the learned flow for a practical application, finding the maximum a posteriori (MAP) value of a noisy observation. The potential gain of the method is that observations from galaxy survey telescopes or intensity mapping could be de-noised using a normalizing flow to ultimately reach better cosmological constraints. While a simulation driven prior comes with some baryonic uncertainty, the learned prior only needs to be closer to reality than the usual Gaussian assumption to successfully improve de-noising.

In cosmology, apart from our previous work \cite{rouhiainen2021}, flows have recently been used to represent the matter distribution of the universe in \cite{trenf}. This paper designed a rotation equivariant flow, TRENF, specifically for cosmology, while here we used a classical real NVP flow \cite{2016arXiv160508803D} which is only translationally symmetric. The TRENF paper is using flows to measure cosmological parameters, while here we use it to reconstruct a field from a noisy observation. The closest existing works which we are aware of are \cite{score_matching_2020} and \cite{score_matching_2022}, which also aim to improve the posterior of a noisy observation of a Gaussian field, by using a learned prior. However in their case, a score matching approach was used which learns gradients, rather than a normalizing flow that gives the complete normalized PDF. Normalizing flows have the advantage that they are straightforward to use both for sampling and for inference, which makes them easy to interpret and visualize. In the present work we consider the somewhat idealized case of an observation of the matter distribution corrupted by Gaussian noise, and study how much signal to noise can be gained depending on the noise in the experiment. We will use the simulated matter distribution as a proxy for observable non-Gaussian fields that depend on the matter distribution, including the smoothed galaxy field in a galaxy survey, the convergence map of galaxy surveys, or the secondary anisotropies of a high-resolution CMB survey (such as kSZ and CMB lensing) \cite{Carlstrom_2002} \cite{Lewis:2006fu}. Applications to realistic observations will be presented in future work.

In Section~\ref{method}, we outline our normalizing flow architecture, and describe how to find posterior reconstructed maps using a learned prior distribution. We show our MAP results in Section~\ref{MAP_results} for reconstructing noisy N-body simulation data, comparing the flow results with Wiener filtering. In Section~\ref{HMC}, Hamiltonian Monte Carlo is used to explore the parameter space of solutions to the de-noising problem. Finally in Section~\ref{patching}, we demonstrate a method of seamlessly patching together many $128$~px length posterior maps to reconstruct a much larger $1024$~px map that might otherwise be too computationally difficult to address without patching.

The code to reproduce our results is available at \texttt{https://github.com/adamrouhiainen/}\\\texttt{denoising-nf}.


\section{Method}
\label{method}

In this section we describe the normalizing flow as well as the optimization procedure we use to denoise the observed matter field.

\subsection{Flow architecture}

The real NVP flow \cite{2016arXiv160508803D}, used in this work to learn the matter distribution from simulations, is widely used and is expressive and fast for both sampling and inference. Here we outline the network we use, with a more thorough introduction to normalizing flows and the details of our network in Appendix~\ref{appendix_realnvp}. Our implementation resembles \cite{Albergo:2021vyo}, stacking $16$ affine coupling layers \cite{2014arXiv14108516D}, each with their own CNN. We use 3 convolutional layers with kernel size 3 and leaky ReLU activation functions, and we use $12$ hidden feature maps after each of the first and second convolutions. This architecture has a receptive field of $97$~px, which on our data corresponds to a Fourier mode of $k=1.6\times10^{-2}\ h/\mathrm{Mpc}$. In practice however, the real NVP network is learning the larger $k$ modes more accurately \cite{rouhiainen2021}. This is acceptable for our purposes, as we will reconstruct the small $k$ modes with Wiener filtering. Testing different sets of hyperparameters, we found improved generalization to out-of-distribution (OOD) data with this setup of a relatively small 26,336 parameters, at the cost of a larger receptive field.

The real NVP flow can be trained on either periodic or non-periodic data simply by setting the padding mode of the network convolutions to either periodic or zero padding. We use the same flow architecture for periodic data (Section~\ref{MAP_results} and Section~\ref{HMC}) and non-periodic data (Section~\ref{patching}) in this work, only changing the convolution padding mode.

We decided against using a more complex flow architecture such as the Glow normalizing flow~\cite{2018arXiv180703039K}, which normally uses more training parameters than real NVP with the benefit of correlating longer length scales with pixel squeezing and channel mixing operations. We also experimented with the rotationally equivariant TRENF flow \cite{trenf}, which uses $\mathcal{O}\left(100\right)$ number of parameters in a typical setup; however with our implementation we were not able match the real NVP flow results.

\subsection{Finding the posterior with an optimizer}

In a cosmological experiment, such as a survey of the galaxy distribution, one can often assume that the vector of data $d=s+n$ is a sum of mutually uncorrelated signal $s$ and noise $n$. The first step in cosmological data analysis is often to find the MAP $\hat{s}$ of the signal given the data $d$. The MAP is given by maximizing the posterior
\begin{align}
\ln{P(s|d)}&\propto \ln{P(d|s)}+\ln{P(s)}\\
&=-\frac{1}{2}\left(s-d\right)^\mathrm{T}N^{-1}\left(s-d\right)+\ln{P(s)}
\end{align}
with respect to the signal $s$ to find the MAP $\hat{s}$. Here we assumed that the noise of the experiment, which appears in the likelihood, is Gaussian with covariance $N=\langle nn^\mathrm{T}\rangle$, which is usually the case in practice. If we also assume that the signal is a Gaussian field with covariance $S=\langle ss^\mathrm{T}\rangle$, i.e.
that the prior is 
\begin{align}
\ln{P(s)} 
&\propto -\frac{1}{2}s^\mathrm{T}S^{-1}s,
\end{align}
then there is an analytic solution to the maximization, called Wiener filtering, given by
\begin{equation}
    \hat{s}_\text{WF}=S\left(S+N\right)^{-1}d.
\end{equation}

Wiener filtering is very common in cosmology, see for example \cite{Carron:2017mqf}, \cite{Munchmeyer:2019kng}. However, upcoming surveys in cosmology such as Rubin Observatory \cite{LSSTSciBook} or Simons Observatory \cite{simons_observatory} probe the matter distribution with such high resolution, that scales are being measured where the Gaussianity assumption of the signal prior does not hold at all. Until recently, it would have been difficult to improve upon this assumption, because no good analytic expressions for the matter distribution~$P(s)$ at non-Gaussian scales exist. In this work, we introduce the reconstruction of non-Gaussian signal maps $s$ where the prior $\ln{P(s)}$ is learned with a normalizing flow. 

Based on the learned differentiable prior represented by the normalizing flow, we can either find the MAP solution to the posterior $\ln{P(s|d)}$, or perform Hamiltonian Monte Carlo to make probabilistic instances of the posterior $\ln{P(s|d)}$. In this work we demonstrate both applications, but first focus on the MAP for simplicity. A benefit of using a learned prior over directly training a neural network for de-noising is that the noise matrix $N$ only appears in the likelihood term $\ln{P(d|s)}$, which is easy to compute for Gaussian noise. Therefore a single trained flow used as the prior $\ln{P(s)}$ may be used to de-noise any amount of noise $N$.

\section{Results}
\label{MAP_results}

We use the particle mesh code \texttt{FastPM} \cite{Feng:2016yqz} to generate an ensemble of simulations of the matter distribution of patches of the universe, and project them to 2 dimensions for computational simplicity. Using cosmological parameters $\Omega_\mathrm{M}=0.315$ and $\sigma_8=0.811$, we simulate $128^3$ particles in a $512\ \mathrm{Mpc}/h$ side-length periodic box, and run 10 steps from scale factor $a=10$ to $a=1$. The particles are then fitted to mesh, creating 3D arrays of $128^3$~px. We make four 2D projections per 3D box by projecting two dimensions by a quarter of the box length along the third dimension. We make a total of 48,000 periodic matter density maps, split 80-10-10 as training, validation, and test sets. Our real NVP flow is trained on the $128^2$~px 2D projections of the simulations with an RTX A4000, using a batch size of 96 with a random rotation and flip given to each map. We minimize the Kullback–Leibler divergence \cite{2019arXiv191202762P} between the flow mapping of a Gaussian noise base distribution and our simulation target distribution with an Adam optimizer of learning rate $10^{-3}$, reduced by half on when the loss plateaus. The loss converges in $\sim10^6$ training cycles, in about 10~hours.

\begin{figure}[h]
    \centering
    \includegraphics[width=0.65\textwidth]{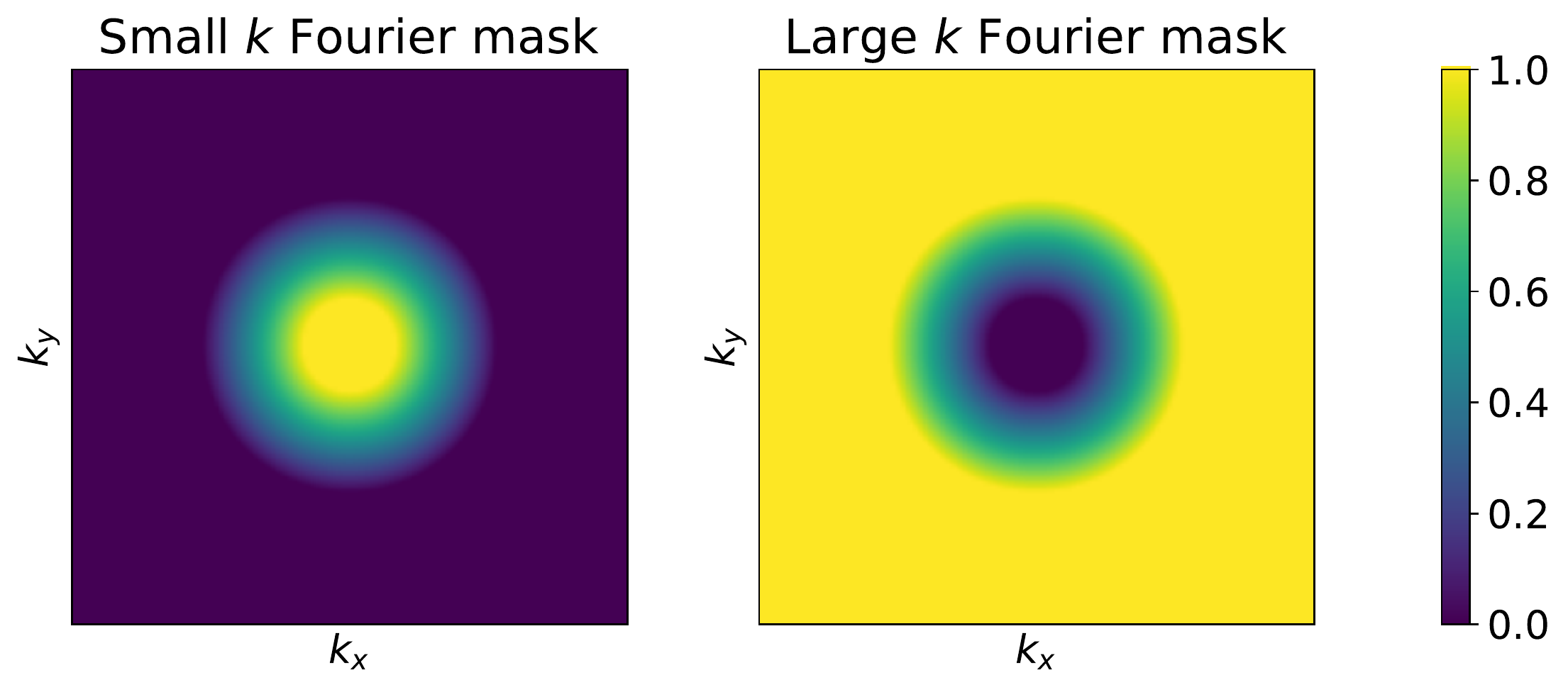}
    \caption{Wiener filtering is accurate on small $k$ modes, and we train a flow to be accurate on large $k$ modes. We apply a small $k$ Fourier mask (left) to Wiener filtered maps, and a large $k$ mask (right) to the flow maps. Our final reconstructed map is the sum in Fourier space of these two results. We found slightly more accurate summary statistics by using a smooth cutoff.}
    \label{fig:fourier_masks}
\end{figure}

After the flow is trained, we make simulated noisy data maps, by adding pixel-wise shot noise to the independent test data, and mask it to mimic a typical survey geometry. On this simulated data, we run an Adam optimizer to find MAP maps with our flow prior by extremizing $\ln{P(s|d)}$. We found that we obtain the best results by using the flow prior only on small non-Gaussian scales, while optimizing the large linear scales with ordinary Wiener filtering (where it is optimal). Thus we make a smooth Fourier cutoff at about $k=0.2\ h/\mathrm{Mpc}$, taking the small $k$ modes from Wiener filtering and the large $k$ modes from the flow posterior. In all results that follow, the "flow" results are calculated after doing this Fourier splitting. An illustration of our Fourier masks is in Fig.~\ref{fig:fourier_masks}.

By splitting off the linear modes, we assume no large-scale to small-scale coupling. This assumption is not exactly correct in cosmology, and our reconstruction is thus not optimal. We will explore modelling such large-scale to small-scale coupling in the future using conditional normalizing flows, potentially improving the reconstruction further by conditioning the flow on the large-scale environment. However, our current implicit factorization assumption of the PDF in Fourier space is sufficient to improve the reconstruction as we shall now see.

Examples demonstrating how our flow posterior reconstructs information on noisy, masked maps are shown in Fig.~\ref{fig:map_maps}. The noise in these examples has st. dev. $0.5\tilde{\sigma}$ and $1.0\tilde{\sigma}$, where we call $\tilde{\sigma}$ the pixel-wise st.~dev. of our training data.

\newcommand\fourimwidth{0.2485}
\newcommand\fourhspace{-2.mm}
\newcommand\fourvspace{-6.5mm}

\begin{figure}[h]
    \centering
    {\Large$0.5\tilde{\sigma}$ noise}
    
    \medskip
    \begin{subfigure}{\fourimwidth\textwidth}
        \includegraphics[width=\textwidth]{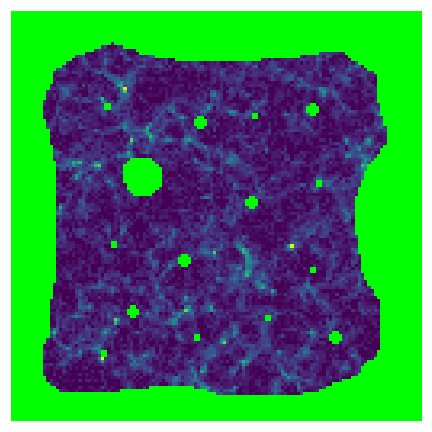}
        \vspace*{\fourvspace}
        \caption*{Observed (noisy, masked)}
    \end{subfigure}
    \hspace*{\fourhspace}
    \begin{subfigure}{\fourimwidth\textwidth}
        \includegraphics[width=\textwidth]{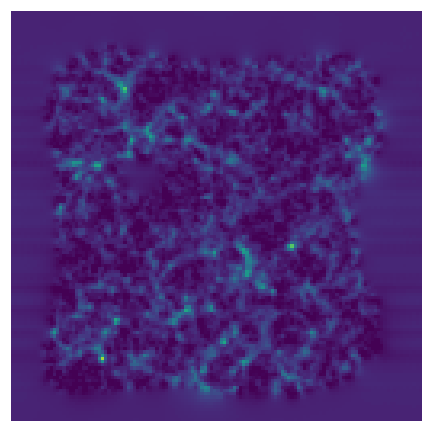}
        \vspace*{\fourvspace}
        \caption*{Wiener filtered}
    \end{subfigure}
    \hspace*{\fourhspace}
    \begin{subfigure}{\fourimwidth\textwidth}
        \includegraphics[width=\textwidth]{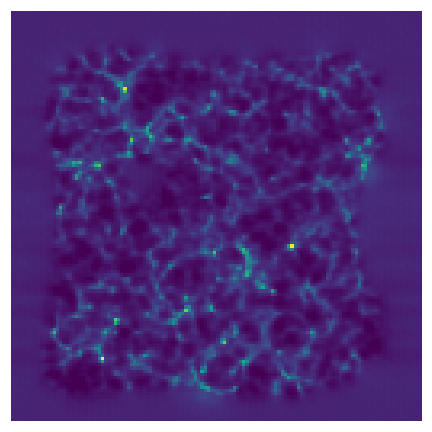}
        \vspace*{\fourvspace}
        \caption*{Flow posterior}
    \end{subfigure}
    \hspace*{\fourhspace}
    \begin{subfigure}{\fourimwidth\textwidth}
        \includegraphics[width=\textwidth]{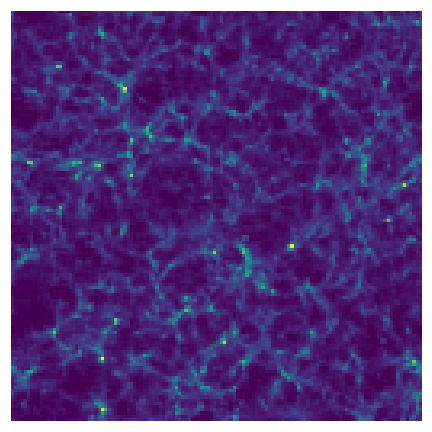}
        \vspace*{\fourvspace}
        \caption*{Truth}
    \end{subfigure}
    
    \medskip
    
    {\Large$1.0\tilde{\sigma}$ noise}
    
    \medskip
    
    \centering
    \begin{subfigure}{\fourimwidth\textwidth}
        \includegraphics[width=\textwidth]{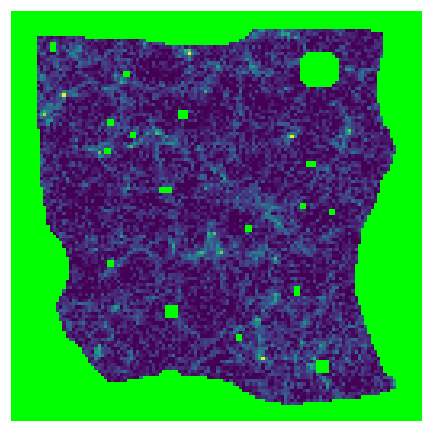}
        \vspace*{\fourvspace}
        \caption*{Observed (noisy, masked)}
    \end{subfigure}
    \hspace*{\fourhspace}
    \begin{subfigure}{\fourimwidth\textwidth}
        \includegraphics[width=\textwidth]{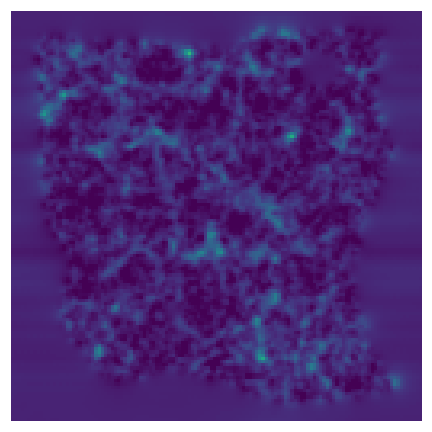}
        \vspace*{\fourvspace}
        \caption*{Wiener filtered}
    \end{subfigure}
    \hspace*{\fourhspace}
    \begin{subfigure}{\fourimwidth\textwidth}
        \includegraphics[width=\textwidth]{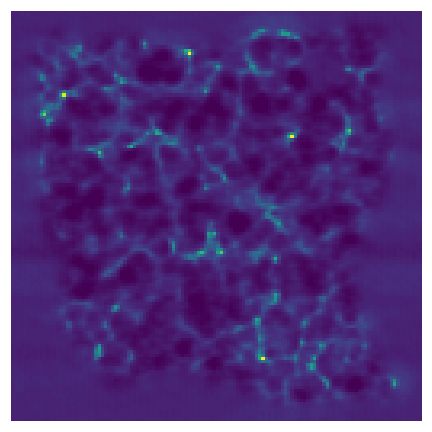}
        \vspace*{\fourvspace}
        \caption*{Flow posterior}
    \end{subfigure}
    \hspace*{\fourhspace}
    \begin{subfigure}{\fourimwidth\textwidth}
        \includegraphics[width=\textwidth]{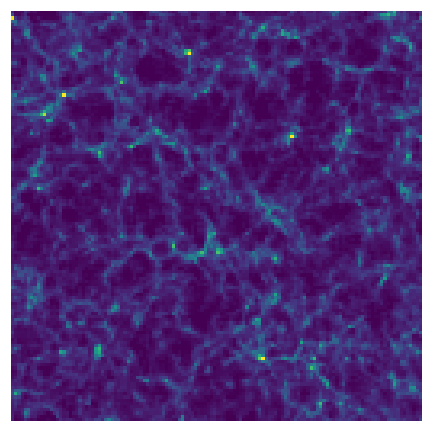}
        \vspace*{\fourvspace}
        \caption*{Truth}
    \end{subfigure}
    
    \caption{Observed, reconstructed posterior, and truth maps for $0.5\tilde{\sigma}$ and $1.0\tilde{\sigma}$ noise and a mask (where $\tilde{\sigma}$ is the st. dev. of the training data). Wiener filtering reduces noise on large and moderate length scales at the cost of over-smoothing small length scales. The flow posterior maps correctly retain the high frequency modes. The maps have length $128$~px, with physical length $512\ \mathrm{Mpc}/h$ and a projected depth of $\Delta z=128\ \mathrm{Mpc}/h$.}
    \label{fig:map_maps}
\end{figure}

We measure the quality of our posterior maps in several ways: the MSE per pixel, the power spectra $P(k)$, and the reconstruction noise defined by
\begin{equation}
    N^\mathrm{rec}(k)=\langle(\varepsilon^\mathrm{rec})^\dagger\varepsilon^\mathrm{rec}\rangle
\end{equation}
where
$\varepsilon^\mathrm{rec}=s^\mathrm{rec}-s^\mathrm{truth}$. We also measure the accuracy of the reconstruction with the Fourier mode cross-correlation coefficient:
\begin{equation}
    r(k)=\frac{P^\mathrm{true,\ rec}(k)}{\sqrt{P^\mathrm{true}(k)P^\mathrm{rec}(k)}}.
\end{equation}
where $P^\mathrm{true,\ rec}(k)$ is the cross power spectrum. By "$\mathrm{rec}$" in these equations, we mean either the flow posterior or the Wiener filtered map.

We present results for reconstructing 100 maps in our test set for the $0.5\tilde{\sigma}$ and $1.0\tilde{\sigma}$ noise setups. Fig.~\ref{fig:map_ps} shows plots of the power spectrum of the posterior maps and $N(k)$ (left), and cross-correlation~(right). We find improvement with the flow MAP against Wiener filtering on all scales above the nonlinear scale $k\sim0.2\ h/\mathrm{Mpc}$, with an improvement of up to a factor of 2 in this setup for large~$k$. For $0.5\tilde{\sigma}$ noise, the flow posterior MSE is 29\% lower than the Wiener filtering MSE, and for $1.0\tilde{\sigma}$ noise the flow MSE is 22\% lower.

The improvement factor also depends strongly on the non-Gaussianity of the field. For example if we reduce the depth range (given by the radial distance $z$) of our 2D maps from $\Delta z=128\ \mathrm{Mpc}/h$ to $\Delta z=32\ \mathrm{Mpc}/h$, we find an improvement on the smallest scales by a factor of about $3.5$ in $r$. It will be interesting to study the possible improvement with volumetric data using a 3D flow.

\newcommand\twoimwidth{0.4999}

\begin{figure}
    \centering
    {\Large$0.5\tilde{\sigma}$ noise}
    
    \medskip
    
    \includegraphics[scale=\twoimwidth]{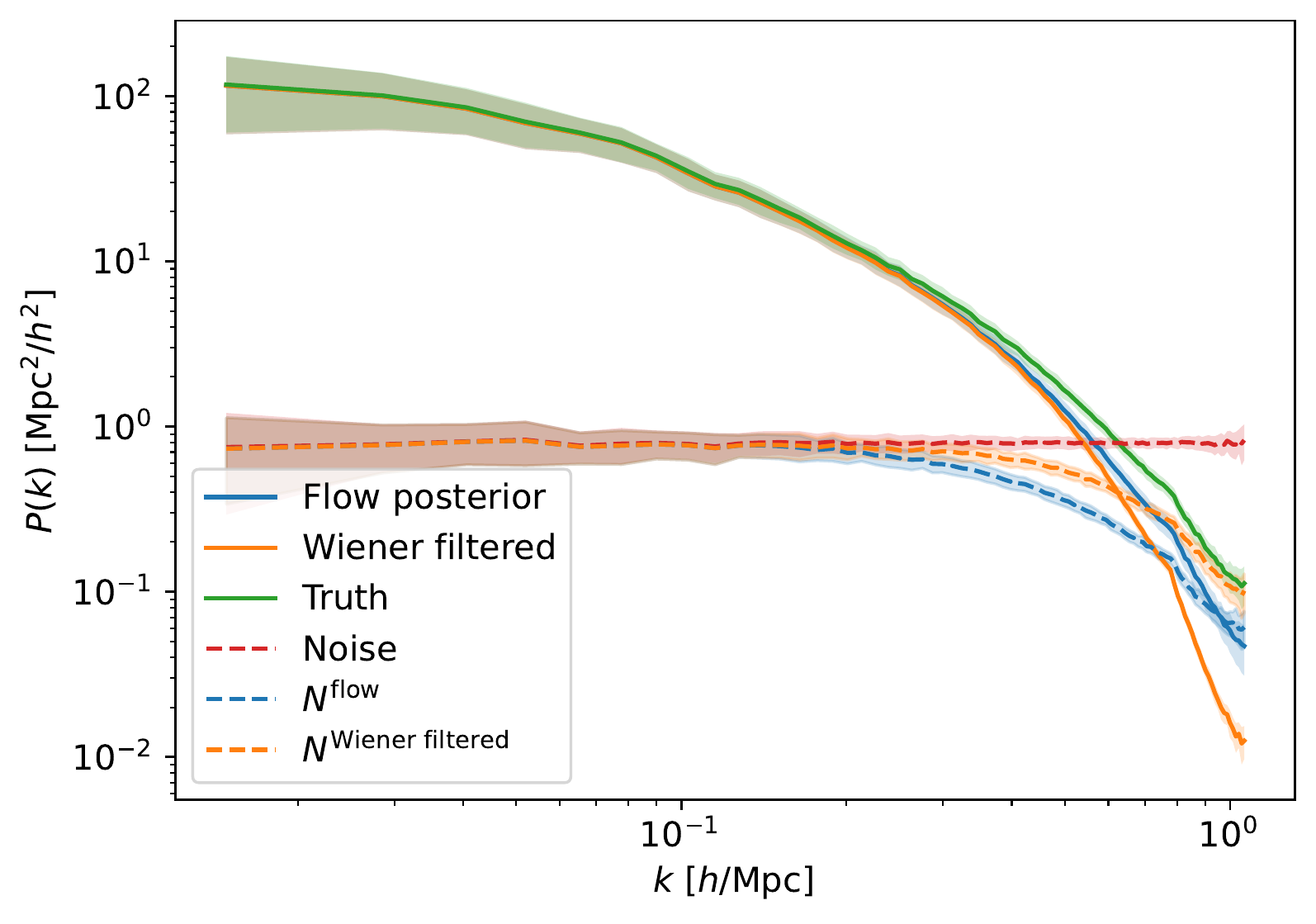}
    \includegraphics[scale=\twoimwidth]{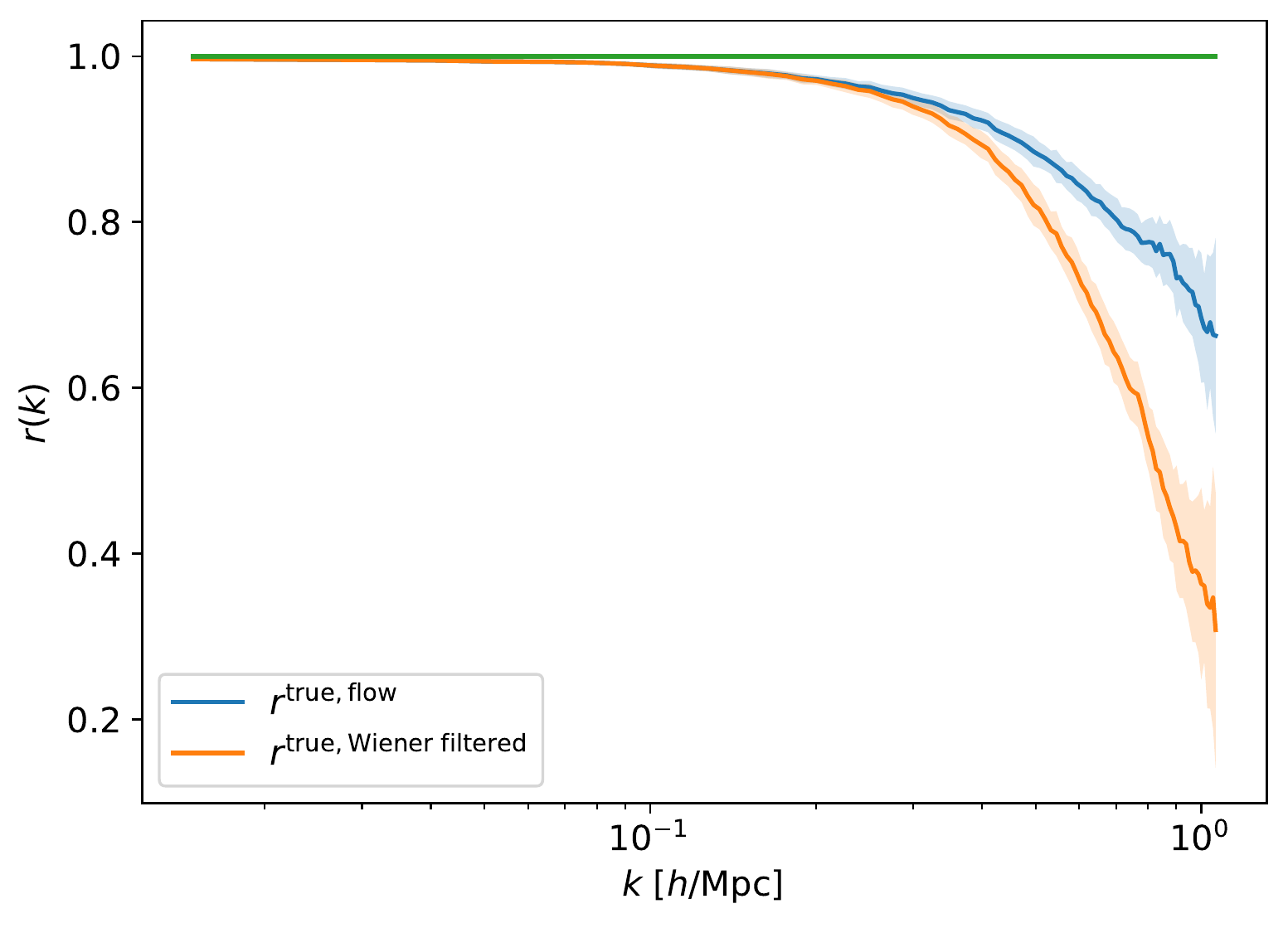}
    
    \medskip
    
    {\Large$1.0\tilde{\sigma}$ noise}
    
    \medskip
    
    \includegraphics[scale=\twoimwidth]{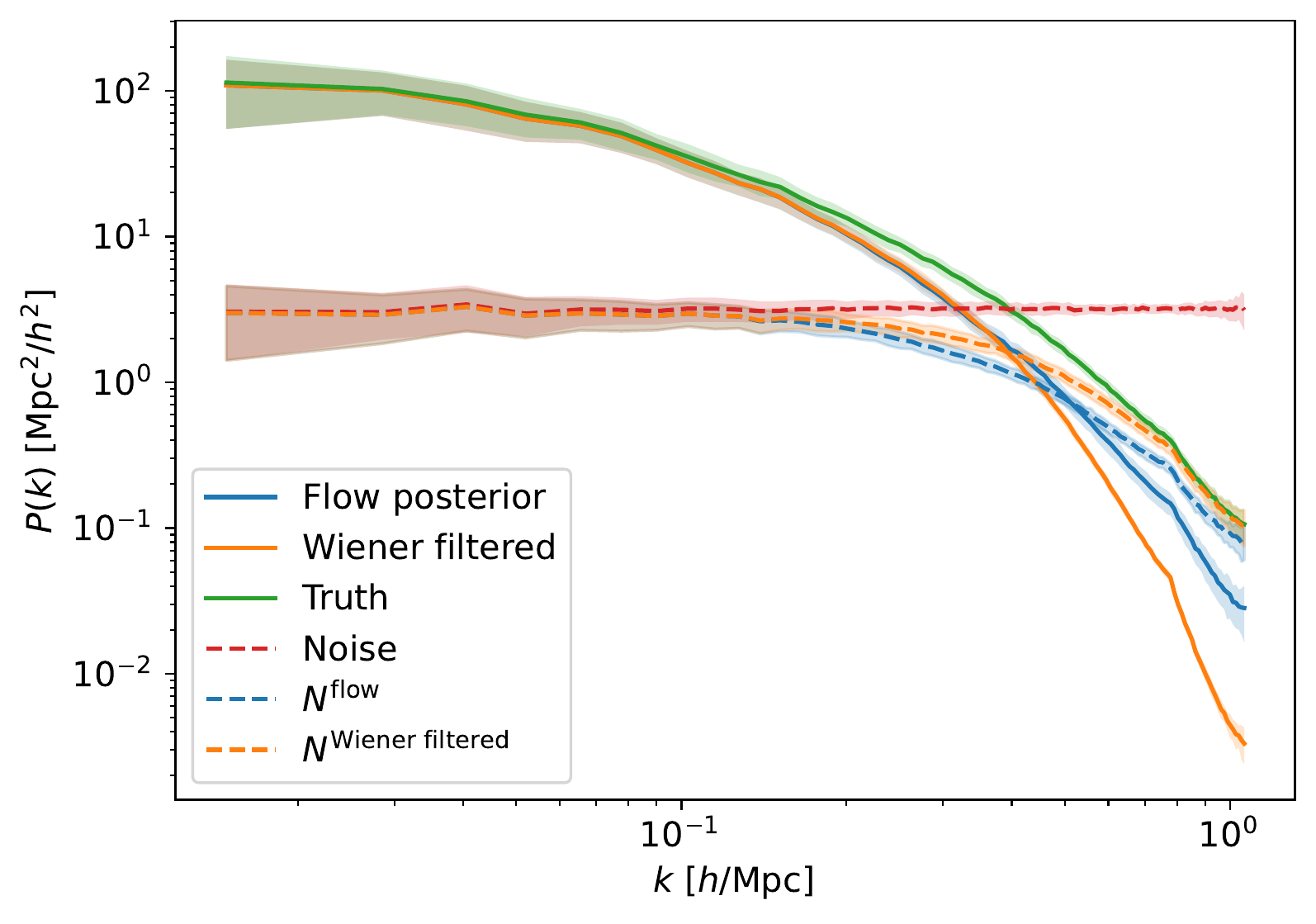}
    \includegraphics[scale=\twoimwidth]{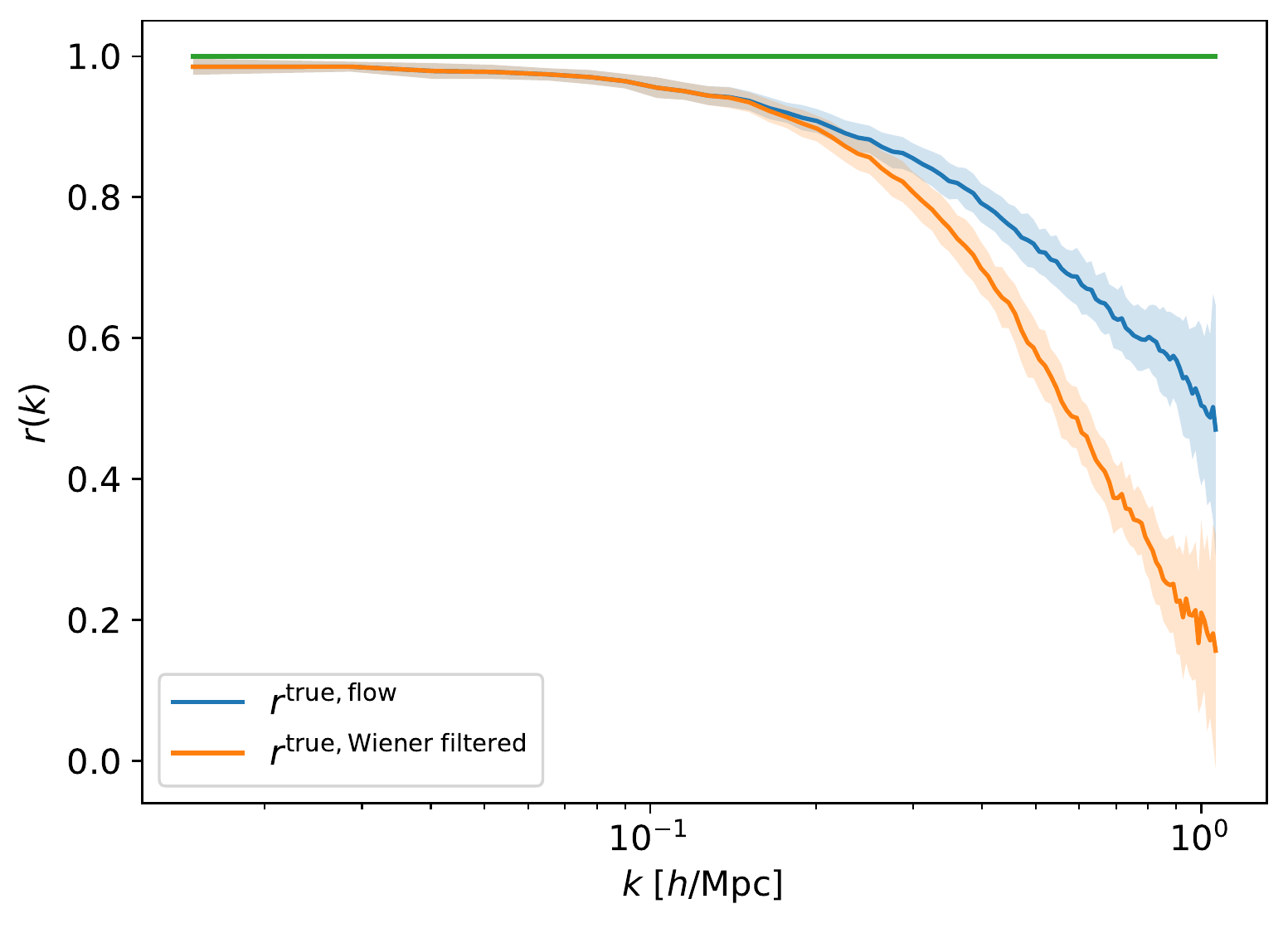}
    \caption{Power spectra (left) and cross-correlation (right) comparing our flow posterior with Wiener filtering for de-noising $0.5\tilde{\sigma}$ and $1.0\tilde{\sigma}$ noise, averaged over 100 maps, along with $1$ st. dev. confidence intervals, computed without a mask. We find improvement with the flow over Wiener filtering on all modes above the nonlinear scale $k\sim0.2\ h/\mathrm{Mpc}$: the flow power spectrum is closer to the truth, $N^\mathrm{flow}$ is lower than $N^\mathrm{Wiener\ filtered}$, and $r^\mathrm{true,\ flow}$ has up to a factor of $2$ improvement at larger $k$ modes. In this example the projected depth of the map is $\Delta z=128\ \mathrm{Mpc}/h$ which is relatively large; a smaller depth leads to more non-Gaussianity and thus even larger improvements.}
    \label{fig:map_ps}
\end{figure}

\begin{figure}[h]
    \centering
    {\includegraphics[scale=\twoimwidth]{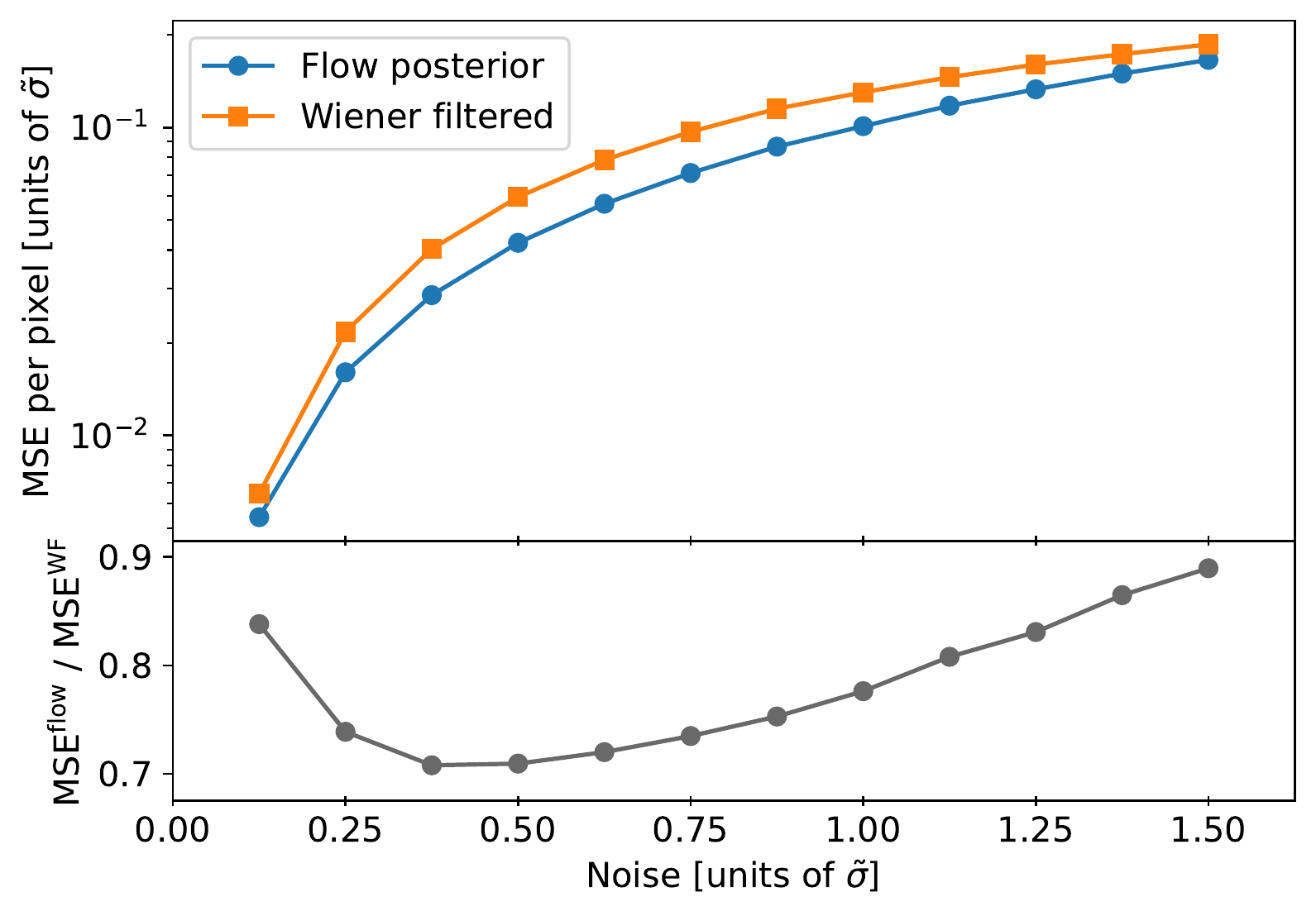}}
    \caption{MSE per pixel between posterior and truth maps for a range of noise levels, calculated on 100 maps at each point. The flow improves de-noising at all noise levels relative to Wiener filtering, with the largest improvement at a noise half the signal for a 30\% MSE reduction over Wiener filtering.}
    \label{fig:mse}
\end{figure}

We also examined how the improvement in the reconstruction depends on the noise in the map (without a mask). Fig.~\ref{fig:mse} shows the MSE per pixel calculated as a function of noise level (relative to~$\tilde{\sigma}$), comparing the flow posterior with Wiener filtering. We find a lower MSE with the flow at all noise levels. The greatest reduction in the flow's MSE over Wiener filtering is for the noise around half of the signal ($0.5\tilde{\sigma}$), giving about a $30\%$ improvement, with lesser improvements at either the low or high noise limit. Intuitively, if the noise is small, the gains by a better prior will be small since the prior matters less. On the other hand, for very large noise, the prior will begin to dominate over the likelihood. A difficulty, as we have found, is that the normalizing flow is trained on IID data, but the posterior optimization enters domains of the configuration space that are OOD and thus the flow may not generalize well to such cases \cite{nf_ood}. We found that it is advantageous for generalization to use a flow with relatively few training parameters, and that splitting the $k$ modes as explained above helps at large noise.



\section{Generating posterior samples with Hamiltonian Monte Carlo}
\label{HMC}

In the previous section we found the MAP solution to the de-noising problem. We can also generate many diverse instances of posterior maps with the Hamiltonian Monte Carlo (HMC) algorithm. HMC obtains samples from the parameter space of a probability density $U\left(s\right)$ containing parameters $s$, and introduces auxiliary momentum variables $p$. The momentum adds a term $\frac{1}{2}p^\mathrm{T}M^{-1}p$ to the Hamiltonian (where the mass matrix $M$ is often taken to be diagonal), allowing the use of Hamiltonian dynamics to explore the space of posterior solutions for $s$. The Hamiltonian is
\begin{align}
    H(s, p)&=U(s)+\frac{1}{2}p^\mathrm{T}M^{-1}p\\
    &=-\ln{P\left(s|d\right)}+\frac{1}{2}p^\mathrm{T}M^{-1}p,
\end{align}
which corresponds to the familiar equations of motion,
\begin{align}
    \frac{\mathrm{d}s}{\mathrm{d}t}&= \frac{\partial H}{\partial p},\\
    \frac{\mathrm{d}p}{\mathrm{d}t}&=-\frac{\partial H}{\partial s}.
\end{align}
The position $s$ and momentum $p$ are evolved under Hamiltonian dynamics for a small time step by numerically integrating Hamilton's equations of motion. Every HMC sample containing the set of parameters $s$ will be correlated to its previous sample in the Monte Carlo chain; the goal is to run a long enough chain to produce a number of uncorrelated samples.

We generate HMC samples with \texttt{hamiltorch} \cite{hamiltorch}, and use the no U-turn sampler (NUTS) \cite{NUTSHMC}. A main feature of NUTS is that it requires less fine-tuning of the step size. Each chain has a burn-in set of $500$ samples to set a step size, where there are $100$ steps per sample. The adapted step size from NUTS is about $0.02\tilde{\sigma}$ with a target acceptance rate of 0.8. For a single noisy map, we generate $2000$ HMC samples in this way, taking about $2$ hours to run each chain.

\begin{figure}[H]
    \centering
    \begin{subfigure}{\fourimwidth\textwidth}
        \scalebox{1}[1]{\includegraphics[width=\textwidth]{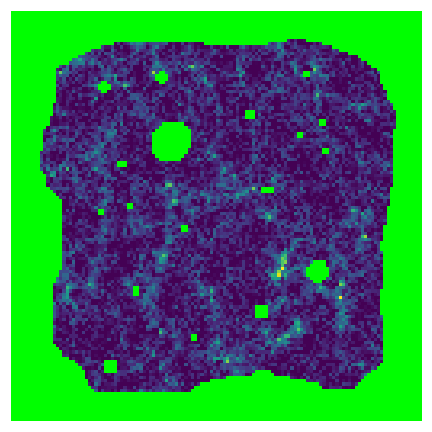}}
        \vspace*{\fourvspace}
        \caption*{Observed (noisy, masked)}
    \end{subfigure}
    \hspace*{\fourhspace}
    \begin{subfigure}{\fourimwidth\textwidth}
        \scalebox{1}[1]{\includegraphics[width=\textwidth]{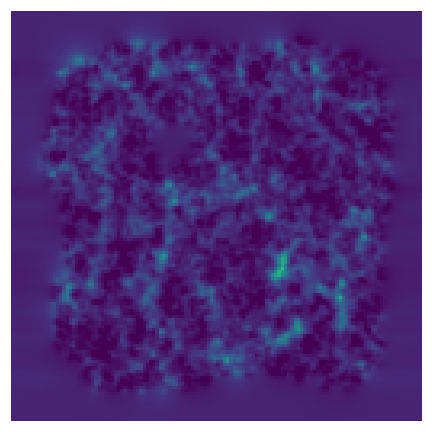}}
        \vspace*{\fourvspace}
        \caption*{Wiener filtered}
    \end{subfigure}
    \hspace*{\fourhspace}
    \begin{subfigure}{\fourimwidth\textwidth}
        \scalebox{1}[1]{\includegraphics[width=\textwidth]{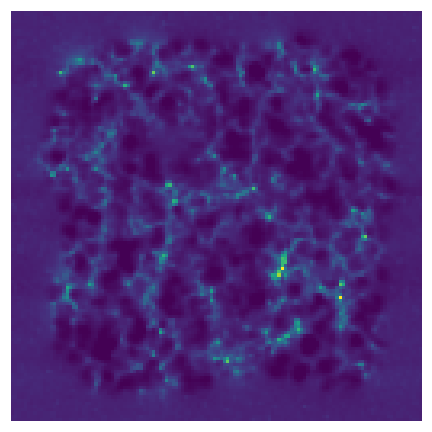}}
        \vspace*{\fourvspace}
        \caption*{Mean of 100 HMC samples}
    \end{subfigure}
    \hspace*{\fourhspace}
    \begin{subfigure}{\fourimwidth\textwidth}
        \scalebox{1}[1]{\includegraphics[width=\textwidth]{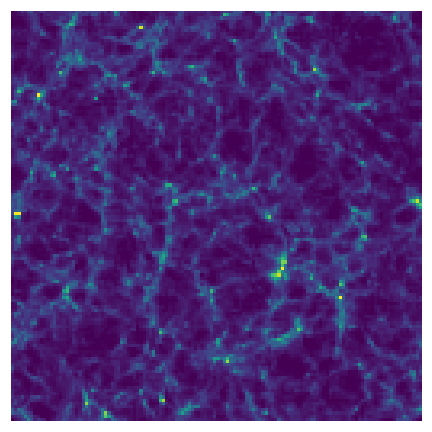}}
        \vspace*{\fourvspace}
        \caption*{Truth}
    \end{subfigure}
    
    \medskip
    
    \begin{subfigure}{\fourimwidth\textwidth}
        \scalebox{1}[1]{\includegraphics[width=\textwidth]{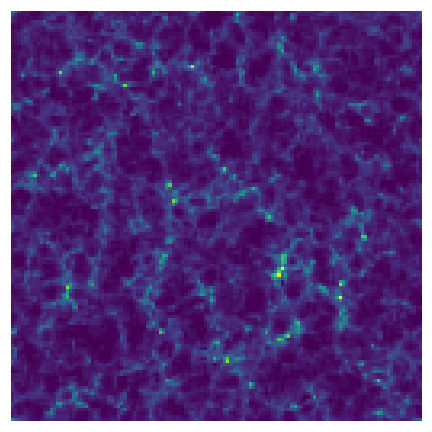}}
        \vspace*{\fourvspace}
        \caption*{HMC sample}
    \end{subfigure}
    \hspace*{\fourhspace}
    \begin{subfigure}{\fourimwidth\textwidth}
        \scalebox{1}[1]{\includegraphics[width=\textwidth]{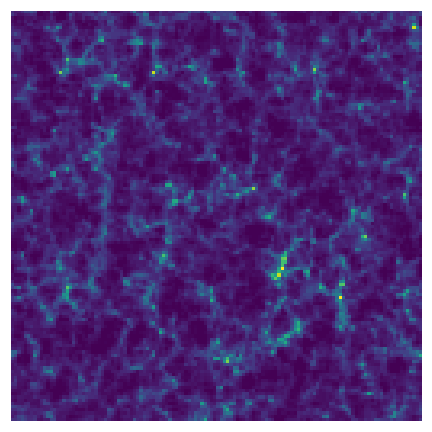}}
        \vspace*{\fourvspace}
        \caption*{HMC sample}
    \end{subfigure}
    \hspace*{\fourhspace}
    \begin{subfigure}{\fourimwidth\textwidth}
        \scalebox{1}[1]{\includegraphics[width=\textwidth]{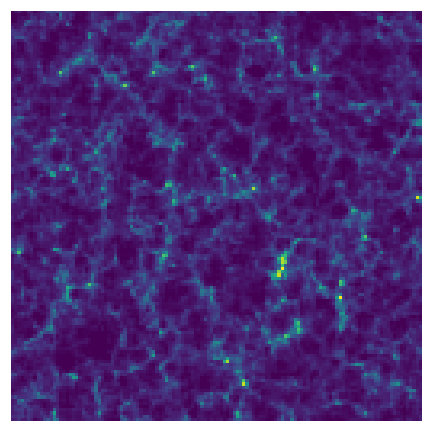}}
        \vspace*{\fourvspace}
        \caption*{HMC sample}
    \end{subfigure}
    \hspace*{\fourhspace}
    \begin{subfigure}{\fourimwidth\textwidth}
        \scalebox{1}[1]{\includegraphics[width=\textwidth]{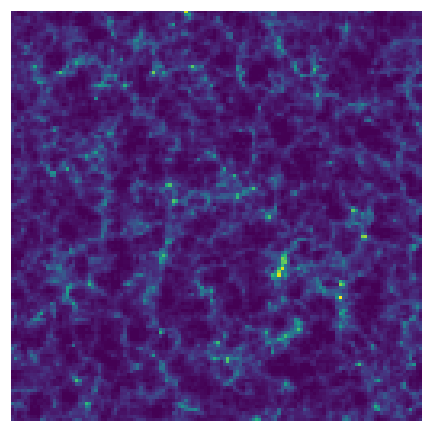}}
        \vspace*{\fourvspace}
        \caption*{HMC sample}
    \end{subfigure}
    
    \caption{(Top row) Observed, reconstructed posterior mean of 100 HMC samples, and truth maps for $1.0\tilde{\sigma}$ noise and a mask. Wiener filtering reduces noise on large and moderate length scales at the cost of over-smoothing small length scales. The HMC posterior mean retains the high frequency modes. (Bottom row) Four HMC samples, showing the variety of possible posterior solutions.}
    \label{fig:hmc_result_128_map}
\end{figure}

Our results for reconstructing a masked map with $1.0\tilde{\sigma}$ noise is in Fig.~\ref{fig:hmc_result_128_map}. We selected $100$ samples evenly spread out along the HMC chain to take a pixel-wise mean as our final posterior mean map. The HMC posterior mean is of similar quality to the MAP solution in the previous section. We also show several HMC samples in Fig.~\ref{fig:hmc_result_128_map}, demonstrating the variety of possible solutions to our de-noising problem.

To obtain the posterior mean (rather than individual samples), we again Wiener filter the large scales and combine them with the HMC samples as described in the previous section. The computational cost of producing uncorrelated HMC samples is lowered with the Fourier splitting, as only the small-scale modes need to be de-correlated in the HMC chain. As is shown in \cite{reconstructVBS}, the autocorrelation length of HMC samples is about $2$ times larger on large length scales.

\begin{figure}[H]
    \centering
    \includegraphics[scale=\twoimwidth]{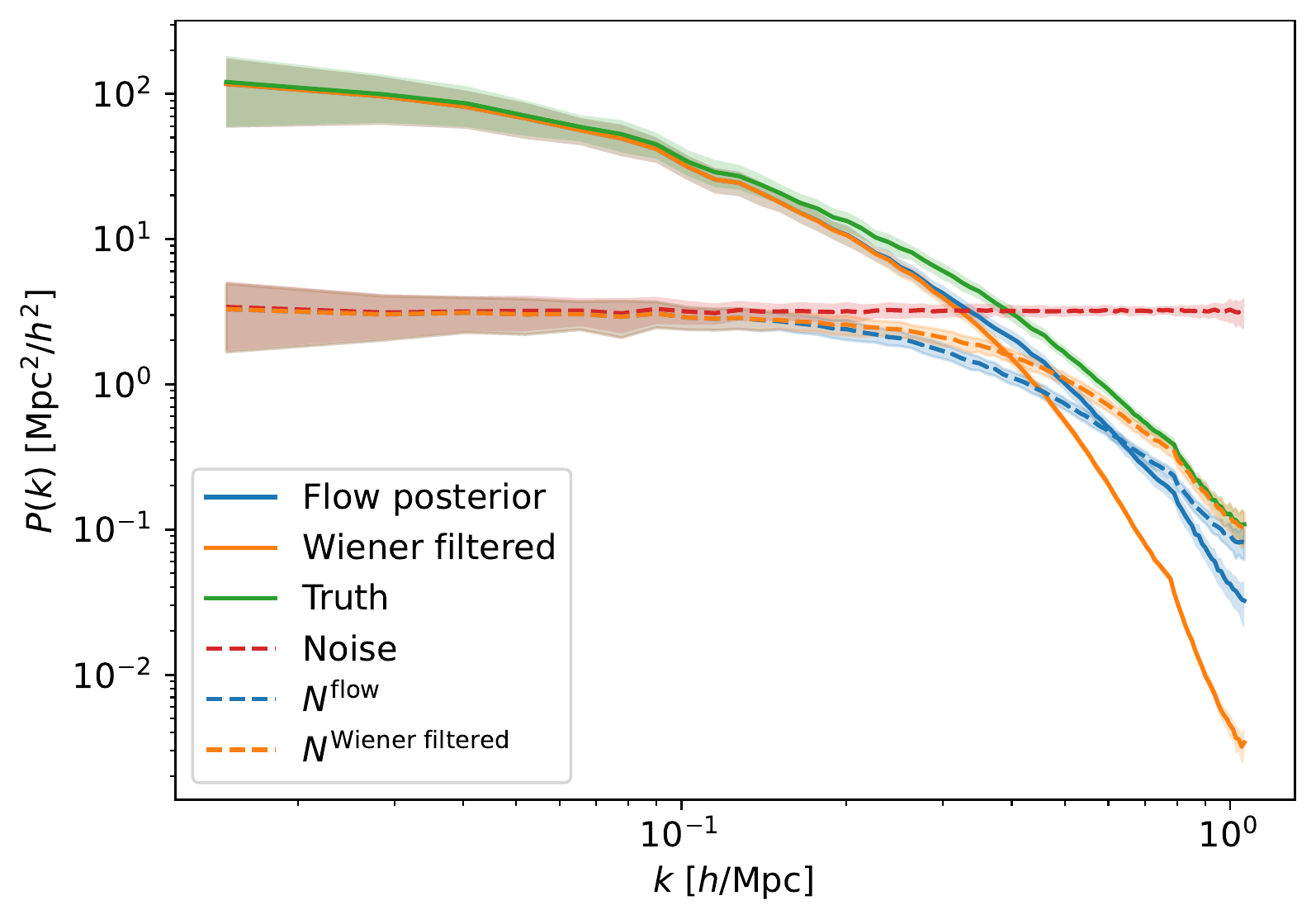}
    \includegraphics[scale=\twoimwidth]{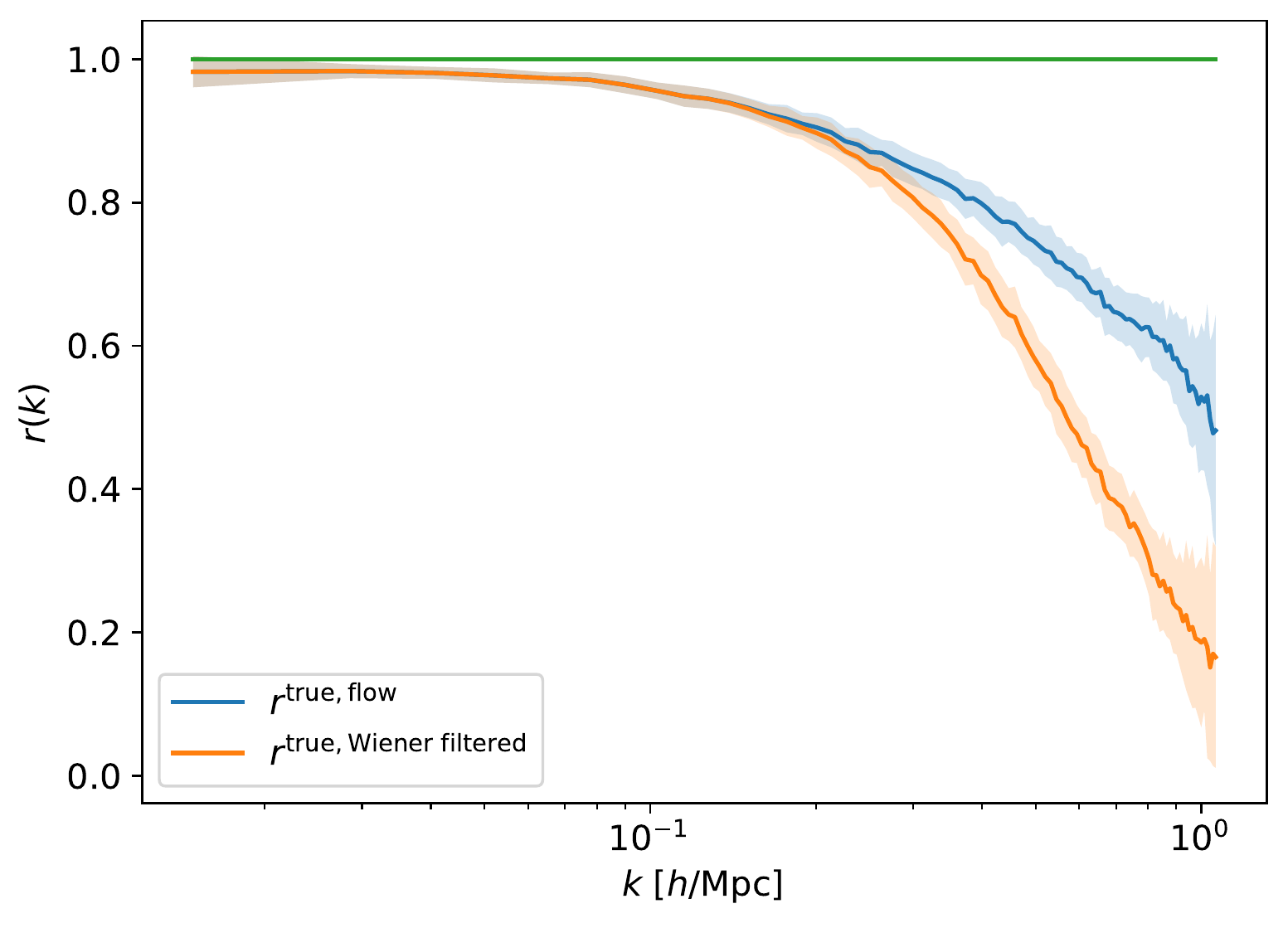}
    \caption{Power spectra (left) and cross-correlation (right) comparing our HMC flow posterior mean with Wiener filtering for de-noising $1.0\tilde{\sigma}$ noise. The posterior mean results are similar to the MAP results in Fig.~\ref{fig:map_ps}. Results are averaged over 100 posterior mean maps, each of which is computed from $100$ HMC samples.}
    \label{fig:hmc_ps_128}
\end{figure}

We also show summary statistics for running HMC chains on 100 different $1.0\tilde{\sigma}$ noisy maps (without a mask now). In Fig.~\ref{fig:hmc_ps_128} are the power spectra (left) and cross-correlation (right), averaged over the 100 different posterior mean maps (where each posterior mean is averaged over 100 HMC samples). The summary statistics here are similar to the previous section, except that the power spectrum of the posterior mean is a bit closer to the truth than the MAP solution.


While the posterior mean HMC maps lose power at large $k$ due to averaging over diverse large $k$ features between samples, the individual HMC samples do have the correct power spectrum. As we showed in \cite{rouhiainen2021}, normalizing flows are capable of generating matter distributions that match their training set power spectrum and non-Gaussianity nearly identically. Making samples from the posterior is required for a fully Bayesian analysis such as, for example, the CMB lensing reconstruction in \cite{Millea:2020iuw}.


\section{Patching maps together to reconstruct large maps}
\label{patching}

Above we discussed that we use an ordinary Wiener filter on large, linear scales, and only use the flow on small nonlinear scales. This Fourier splitting also allows us to reconstruct very large maps with a patching procedure, without the need to increase the flow dimension. Wiener filtering is computationally more tractable, so we may Wiener filter an entire large map, while the computationally more difficult flow posterior is computed on smaller patches in parallel. By patching the smaller flow filtered maps onto the large Wiener filtered map, we avoid having to train an extremely large flow.

As an example, we reconstruct a large map of length $n_\text{L}=1024$~px with a flow trained on non-periodic small maps of length $n_\text{S}=128$~px, but we can patch together arbitrarily large maps in principle. We use the same network architecture as described in Section~\ref{introduction}, with zeros for the convolution padding to respect the non-periodicity. Our training data is as described in Section~\ref{MAP_results}, except our $128$~px 2D projections are cut out of $384$~px 3D simulations to get non-periodicity.

The reconstruction and patching procedure is as follows.
\begin{enumerate}
    \item Divide the large, $n_\text{L}$ length periodic map into a number of $(2n_\text{L}/n_\text{S})^2$ evenly spaced small maps of length $n_\text{S}$. These small maps have an $n_\text{S}\times n_\text{S}/2$ overlapping region with each of their four neighboring maps.
    \item Reconstruct the small-scale modes in these non-periodic small maps with a trained flow.
    \item Reconstruct the large-scale modes by applying Wiener filtering to the entire large map.
    \item To avoid discontinuities near the edges of the small maps, add only the the large $k$ modes from only the center $n_\text{S}/2$ length square region of the flow reconstructed maps to the Wiener filtered large map.
\end{enumerate}

The critical step that allows the smaller maps to be patched together without discontinuities is the Wiener filtering applied to the entire large map. This large-scale Wiener filtering is well within our computational constraints for very large maps encountered in cosmology, while training a flow on very large maps may be computationally infeasible. As explained above, by Wiener filtering large scales, we implicitly assume a factorization of the PDF in Fourier space. Here, by patching, we also assume a factorization in real space on the scale of the patches. Neither of these factorizations hold exactly true in cosmology, but they are sufficient approximations for our goal to to improve the reconstruction.

\begin{figure}[H]
    \centering
    \begin{subfigure}{0.63\textwidth}
        \includegraphics[trim={2mm 2mm 2mm 2mm}, clip, width=\textwidth]{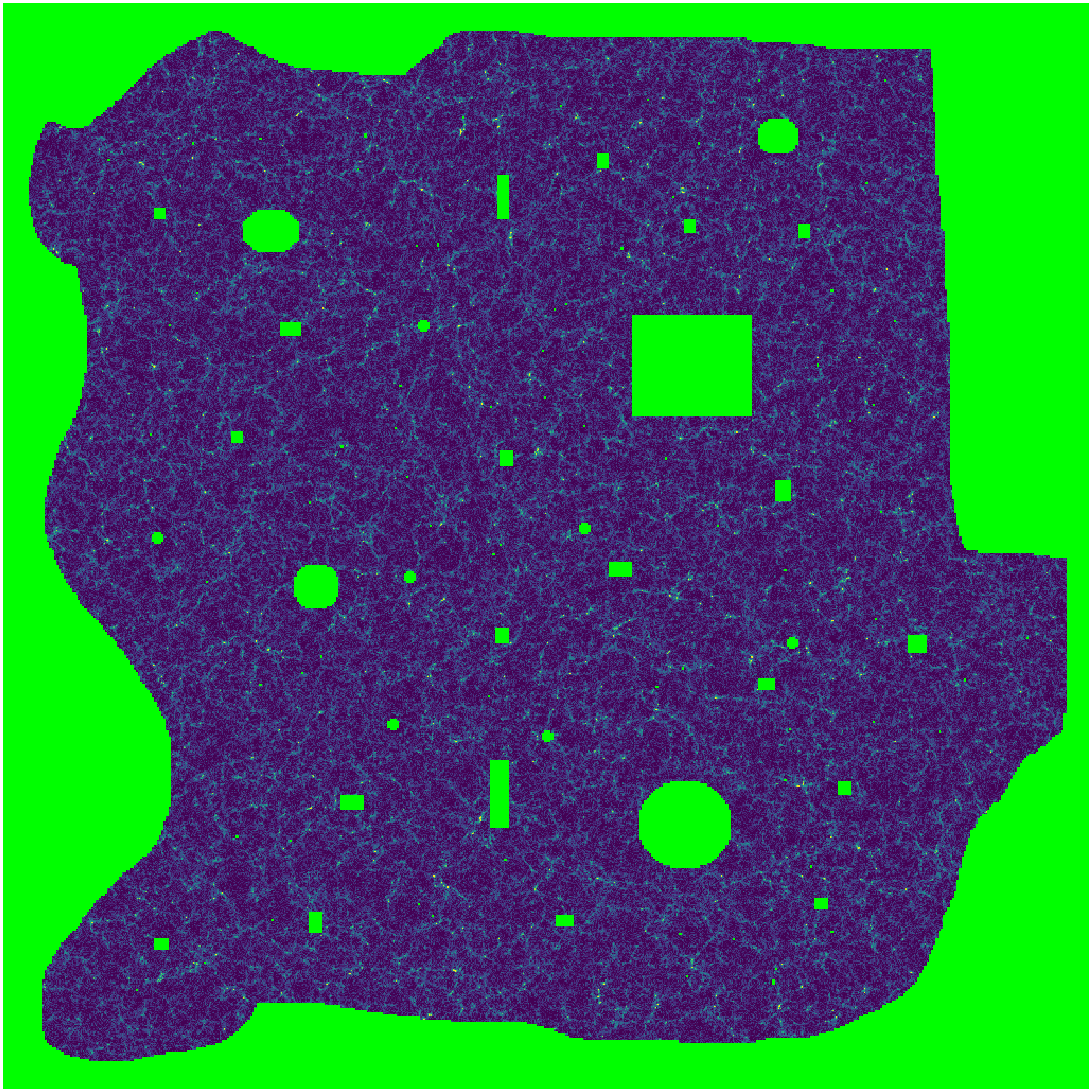}
        \vspace*{\fourvspace}
        \caption{Observed (noisy, masked)}
    \end{subfigure}
    \begin{subfigure}{0.63\textwidth}
        \includegraphics[trim={2mm 2mm 2mm 2mm}, clip, width=\textwidth]{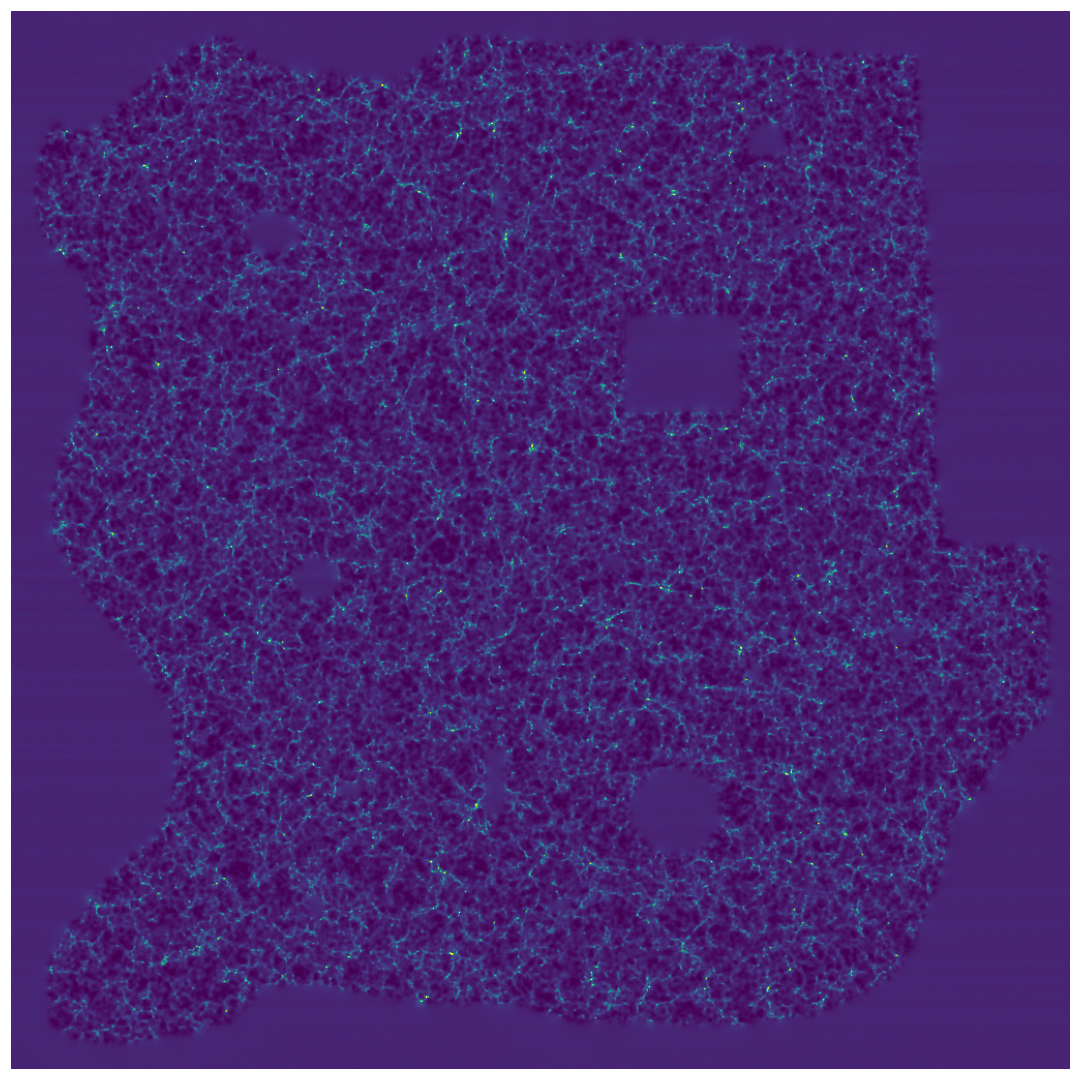}
        \vspace*{\fourvspace}
        \caption{Flow posterior}
    \end{subfigure}
    
    \caption{Observed and reconstructed maps of length $1024$~px, and physical length $4096\ \mathrm{Mpc}/h$ with projected depth $\Delta z=128\ \mathrm{Mpc}/h$. The flow posterior map was patched together with 256 posterior maps of length $128$~px as described in Section~\ref{patching}.}
    \label{fig:result_bigmap}
\end{figure}

\begin{figure}[H]
    \centering
    \includegraphics[scale=\twoimwidth]{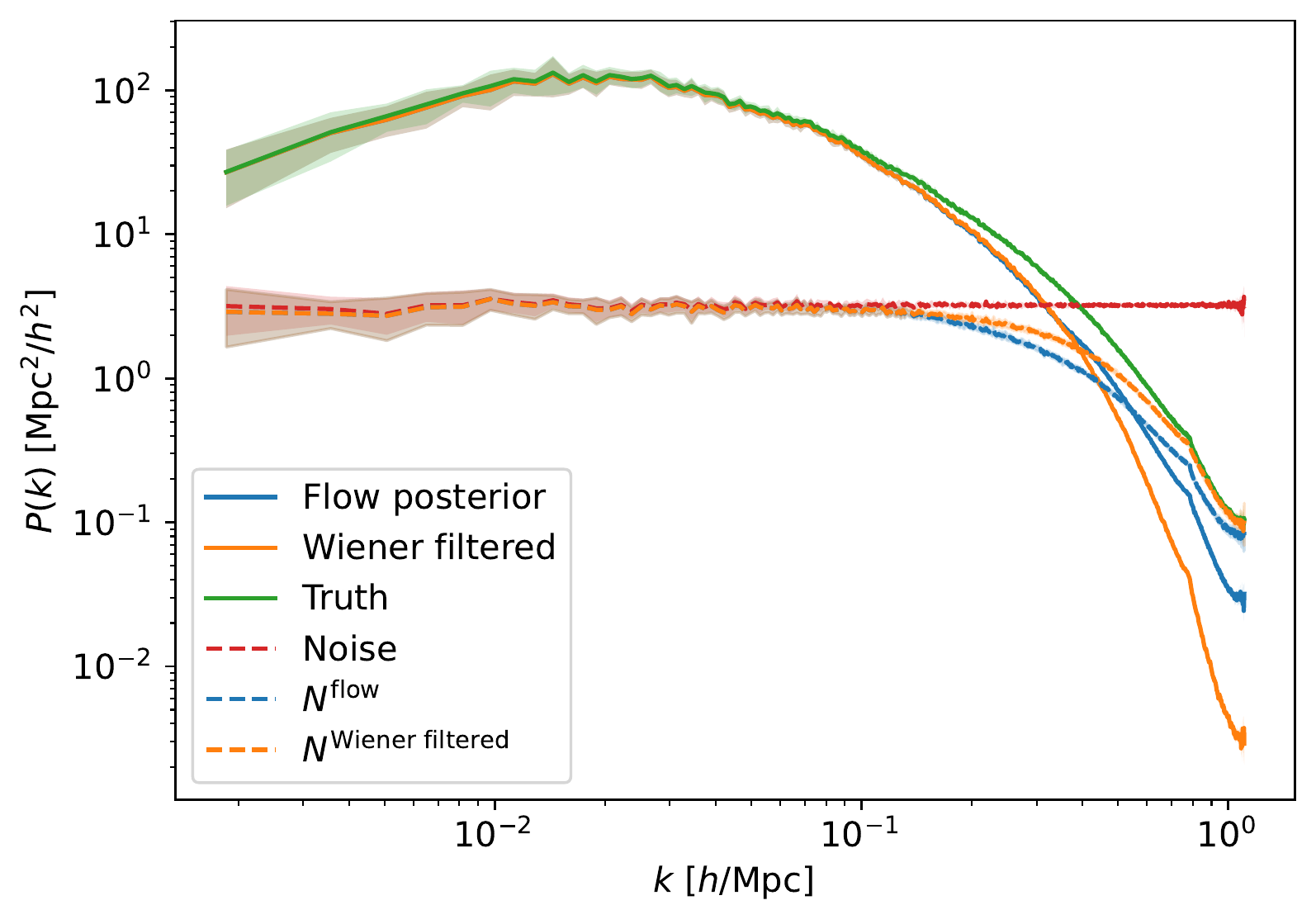}
    \includegraphics[scale=\twoimwidth]{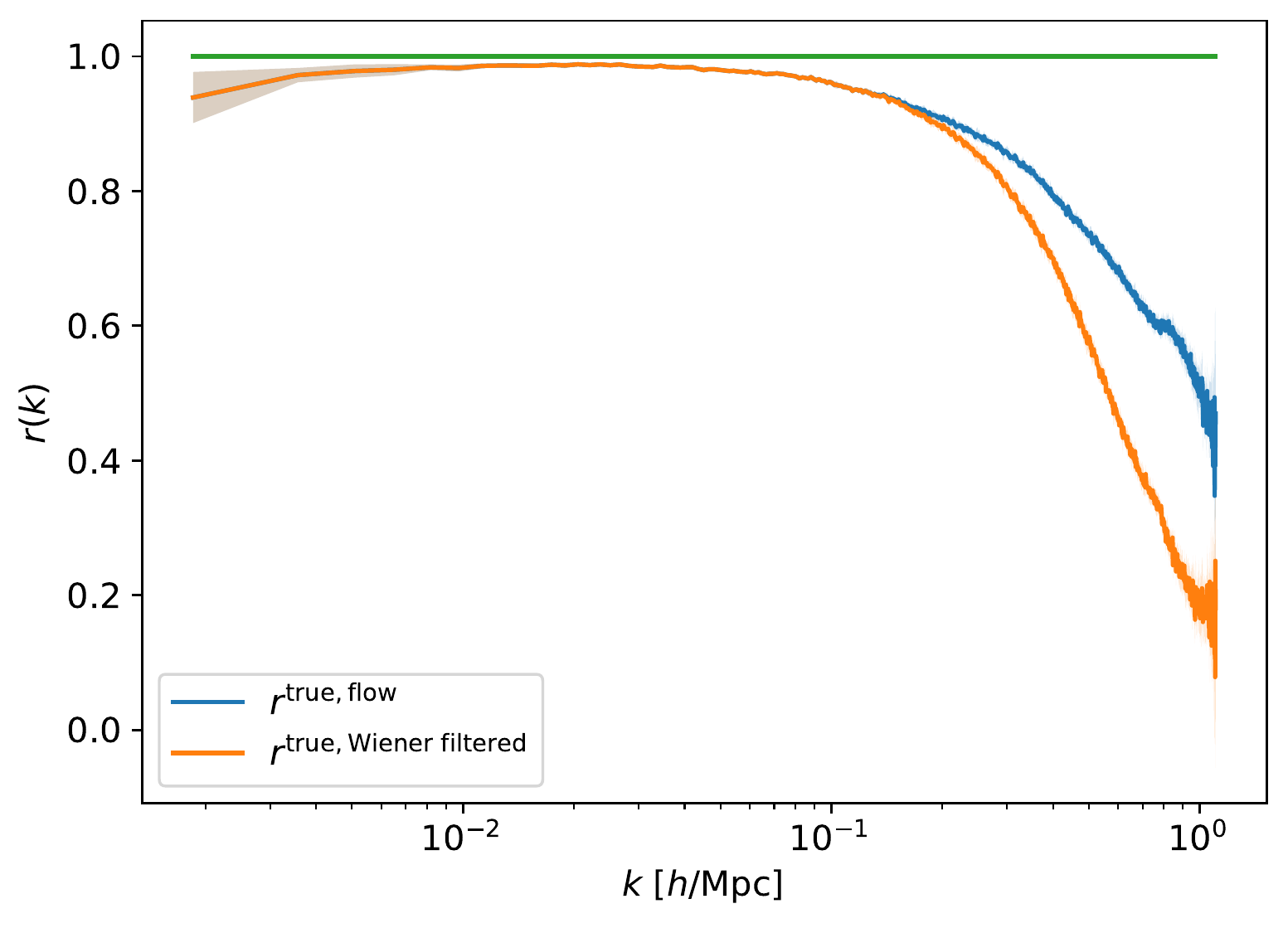}
    \caption{Power spectra (left) and cross-correlation (right) for de-noising $1.0\tilde{\sigma}$ noise $1024$~px maps. We see no evidence of patching in the form of resonances or discontinuities in any of the summary statistics. Results are averaged over 10 different $1024$~px maps, computed without a mask.}
    \label{fig:bigmap_ps_128}
\end{figure}

High-resolution $1.0\tilde{\sigma}$ noise observed and reconstructed $1024$~px maps are shown in Fig.~\ref{fig:result_bigmap}, where the posterior map is reconstructed from 256 small $128$~px maps. There is no visible remnant of a grid where maps were patched together. Additionally, the summary statistics shown in Fig.~\ref{fig:bigmap_ps_128}, which have been averaged over 10 different $1024$~px maps, have no resonances or other oddities giving evidence of the patching.

Our patching method can also aid in generating large HMC samples. The number of steps to reach nearly independent HMC samples of $N$ parameters grows as $\mathcal{O}\left(N^{5/4}\right)$ \cite{neal_mcmc}, so there is a benefit to breaking up a large map and computing HMC samples on individual small maps in parallel, however at the cost of the approximations we just explained. It would be interesting to generalize our approach to a conditional patching to relax these approximations.

\section{Conclusion}

Normalizing flows are a powerful tool to deal with the non-Gaussianity of high-resolution cosmological surveys. The key feature of flows is that their exact likelihoods are tractable, and here we have made use of this feature to de-noise simulated cosmological data. In our example setup, at non-linear scales, we gain up to 30\% reduction in MSE, and a factor of 2 increase in the cross-correlation coefficient, relative to Wiener filtering. The possible improvement depends strongly on the non-Gaussianity of the field. For example, reducing the projected depth from $\Delta z=128\ \mathrm{Mpc}/h$ to $\Delta z=32\ \mathrm{Mpc}/h$, we find an improvement on the smallest scales by a factor of $3.5$ in the cross-correlation. We also demonstrate a method of patching together many small posterior maps to de-noise large maps that would be computationally difficult to address without patching. Our results were obtained by projected 2 dimensional maps. The generalization to 3 dimensions is mathematically straightforward but may pose computational challenges. We will address this in upcoming work.

There are several other interesting follow-up projects. First, we have not included large-scale to small-scale coupling. In ongoing work, we are training a conditional normalizing flow to sample small-scale structure conditioned on the large-scale environment, with an application to super-resolution emulation. Running a conditional flow in inference mode may allow us to improve our reconstruction further. We have also assumed that noise is Gaussian. This approximates instrument noise as well as shot noise, but is not always a good description \cite{2009PhRvL.103i1303S} (e.g. for low mass galaxies with a high number density). 

Furthermore, we would like to use our method to improve constraints on cosmological or astrophysical parameters. As we have seen, our method improves the reconstruction on non-linear scales, starting roughly at the scales where one usually cuts off a cosmological N-point function analysis, such as the power spectrum, due to theoretically uncontrolled non-linearities. To use the flow reconstruction for N-point function analysis would require a reliable simulation modelling of of these scales. Marginalizing over baryonic feedback is in principle possible with flows by conditioning them on unknown feedback parameters \cite{trenf}, but generating reliable training data is difficult. There are however also situations where a small-scale reconstruction can be used reliably for inference of cosmological parameters. This is the case in particular when small-scale modes can be used to reconstruct large-scale modes. For example, a better reconstruction of the mass field from a galaxy survey leads to an improved template for CMB lensing \cite{Schmittfull:2017ffw} or kSZ cross-correlation \cite{Munchmeyer:2018eey,Smith:2018bpn}. Such cross-correlation can be used for example to probe primordial non-Gaussianity \cite{Munchmeyer:2018eey,Schmittfull:2017ffw}. Unknown small-scale physics then appears in terms of biases on large scales which we can marginalize over. In future work we will examine such applications.  


\section{Acknowledgements}

We thank Kimmy Wu for valuable discussions and initial collaboration. This material is based upon work supported by the U.S. Department of Energy, Office of Science, under Award Number DE-SC0022342. Support for this research was provided by the University of Wisconsin-Madison Office of the Vice Chancellor for Research and Graduate Education with funding from the Wisconsin Alumni Research Foundation.

{
\small

\bibliographystyle{unsrt}
\bibliography{main}
}

\appendix

\section{Appendix: Normalizing flows}
\label{appendix_realnvp}

\subsection{Introduction}

A normalizing flow is a natural way to construct flexible probability distributions by transforming a simple base distribution (often Gaussian) to a more complicated target distribution. This is done by applying a series of learned diffeomorphisms to the base distribution. Given a base distribution $p_u(\bf{u})$ of a random variable $\bf{u}$, the target distribution $p_x(\bf{x})$ is given by
\begin{equation}
    p_x(\mathbf{x}) = p_u(\mathbf{u})\left|\det J_T(\mathbf{u})\right|^{-1}
\end{equation}
where $T$ is the transformation $\mathbf{x} = T(\mathbf{u})$, and $J_T$ is its Jacobian. We can construct a transformation $T$ by composing a finite number of simple transformations $T_k$ as
\begin{equation}
    T = T_{K} \circ \cdots \circ T_{1}.
\end{equation}
These simple transformations must have tractable inverses and tractable Jacobian determinants. They depend on learned parameters and can be parametrized using neural networks. In this way very expressive densities can be constructed. Taking $\mathbf{z}_0=\mathbf{u}$ and $\mathbf{z}_K=\mathbf{x}$, the transformation at each step $k$ is
\begin{align}
    \mathbf{z}_{k} = T_k(\mathbf{z}_{k-1})
\end{align}
and the Jacobian determinant is:
\begin{equation}
\log \left| J_{T}(\mathbf{z})  \right| = \sum_{k=1}^{K} \log \left| J_{T_{k}}(\mathbf{z}_{k-1}) \right|.
\end{equation}

Once the flow is learned, two basic statistical operations can be performed efficiently: density evaluation (what is $p_x(\bf{x})$ given a sample $\bf{x}$) and sampling from $p(\bf{x})$. These operations can be used for statistical inference purposes. The difference between normalizing flows and ordinary neural network techniques are that the former are representing normalized probability densities rather than arbitrary mappings from input to output.

\subsection{Real NVP flow}

The first flow with success in creating high quality images was the real-valued non-volume preserving flow, or \textit{real NVP} flow \cite{2016arXiv160508803D}. This flow is expressive and is fast both for sampling and inference. A simplification of our application compared to real NVP is that the latter was constructed for RGB images (3 channels) while we represent a scalar field (1 channel), so we do not require transformations that mix channels.

\textbf{Affine coupling layer}. The basis of this flow is the affine coupling layer, an operation that rescales and shifts a subset of the random variables depending on the value of the other random variables. The set of all random variables (in our application all pixels of the random field), denoted as $x$, is split into two subsets $x_1$ and $x_2$. Then these parameters are updated as follows:
\begin{align}
    x_1' &= e^{s(x_2)} x_1 + t(x_2) \\
    x_2' &= x_2
\end{align}
Here $s$ and $t$ are vector valued, i.e.\ each pixel in $x_1$ can be rescaled and shifted differently. This transformation guarantees invertibility as well as a triangular Jacobian which is computationally easy to evaluate and invert. While the scaling transformation is simple, its flexibility comes from the free form of the functions $s$ and $t$ and stacking many such layers, with different partitions into subsets.

\textbf{Checkerboard masking}. For images, a standard choice of partition into subsets $x_1$ and $x_2$ is the checkerboard masking proposed in \cite{2016arXiv160508803D}, where every pixel alternates between white and black. On the checkerboard, we either use the white pixels for $x_1$ and the black pixels for $x_2$, or vice versa. For each consecutive affine coupling layer we switch the sets $x_1$ and $x_2$. 

\textbf{CNN to determine the affine parameters}. We now need to define the functions $s(x_2)$ and $t(x_2)$. As $x_2$ are spatially organized, rather than arbitrary collections of random variables, it is natural to use a standard convolutional neural network for this purpose (as was done in \cite{2016arXiv160508803D}) which enforces translational symmetry. The CNN needs to conserve dimension, so we use stride 1 convolutions and no pooling. To implement periodic boundary conditions we use the common approach of circular padding. As in \cite{Albergo:2021vyo} we use 3 convolutional layers with kernel size 3 and leaky ReLU activation functions. The number of channels we use is in this order: 1 (the scalar PDF values), 12 (arbitrary number of feature maps), 12 (arbitrary number of feature maps), 2 (the output variables $s$ and $t$).

\textbf{Stacking the layers}. We stack $K=12$ affine coupling layers, each with their own CNN to parameterize the affine transformation $s$ and $t$.

This architecture has 26,336 trainable parameters. Our PyTorch \cite{NEURIPS2019_9015} implementation of this normalizing flow is taken from \cite{Albergo:2021vyo}, with some modifications.

\subsection{Flow training}

To train the flow, we minimize the forward Kullback–Leibler  (KL) divergence (see Sec.~2.3 of the review \cite{2019arXiv191202762P}), a measure of the relative entropy from the target distribution $p^*_x(\mathbf{x})$ to the base distribution $p_x(\mathbf{x})$. The forward KL divergence can be expressed as
\begin{align}
    \mathcal{L(\phi)}&=D_{\text{KL}}\big(p^*_x(\mathbf{x})\ ||\ p_x\left(\mathbf{x};\phi\right)\big)\\
    &=-\mathbb{E}_{p^*_x(\mathbf{x})}\big(\log p_x(\mathbf{x};\phi)-\log p^*_x(\mathbf{x};\phi)\big)\\
    &=-\mathbb{E}_{p^*_x(\mathbf{x})}\big(\log p_u\left(T^{-1}\left(\mathbf{x};\phi\right)\right)+\log\left|\det{J_{T^{-1}}(\mathbf{x};\phi)}\right|\big)+\mathbb{E}_{p^*_x(\mathbf{x})}\log p^*_x(\mathbf{x};\phi)
\end{align}
where $T$ is the flow transformation, with learned parameters $\phi$. The final term here is a constant that we do not need to calculate. Here \textit{forward} denotes the order of $p^*_x$ and $p_x$ above. The expectation values are estimated as
\begin{equation}\label{eq:loss}
    \mathcal{L}\left(\phi\right)=-\frac{1}{N}\sum_{n=1}^N\left(\log p_u\left(T^{-1}\left(\mathbf{x}_n;\phi\right)\right)+\log\left|\det J_{T^{-1}}(\mathbf{x}_n;\phi)  \right|\right)+\text{const.},
\end{equation}
where $\mathbf{x}_n$ are the training samples from $p^*_x(\mathbf{x})$. Minimizing the Monte Carlo approximation of the KL divergence is thus equivalent to fitting the flow model to the training samples by maximum likelihood estimation.

In this work we use the Adam optimizer to minimize the loss with respect to the parameters $\phi$. We use a learning rate of $10^{-3}$, reduced by half when the loss plateaus using the PyTorch learning rate scheduler \texttt{ReduceLROnPlateau} with patience equal to 10. We train with a batch size of 96.

\end{document}